\documentclass{aa}     

\usepackage{graphicx}
\usepackage{txfonts}

\newcommand{\Mpc}{$h^{-1}$\thinspace Mpc}

\newcommand{\etal}{et al.}

\def\apjl{ApJL}
\def\apjs{ApJS}

\begin{document}

\title{ Groups of galaxies in the SDSS Data Release 5} % \\
      \subtitle{  A group-finder and a  catalogue
}

\author {E.  Tago\inst{1} \and J.  Einasto\inst{1} \and E.  Saar\inst{1}
  \and E.  Tempel\inst{1}  \and M.  Einasto\inst{1} \and J.  Vennik\inst{1}
  \and V.  M\"uller\inst{2}}

\institute{Tartu Observatory, EE-61602 T\~oravere, Estonia
\and 
Astrophysical Institute Potsdam, An der Sternwarte 16, D-14482 Potsdam, Germany
}

\authorrunning{E.  Tago et al. }

\offprints{E.  Tago }

\date{ Received   2007 / Accepted . . .   }

\titlerunning{SDSS DR5 groups}

\abstract{}
{
We extract groups of galaxies from the SDSS Data
Release 5 with the purpose of studying the
supercluster-void network and environmental
properties of groups therein.
Groups of galaxies as density enhancements 
can be used to  determine the luminosity density field
of the network. 
}
{
We use a modified friends-of-friends (FoF) method with
adopted variable linking length in transverse and radial
direction to eliminate selection effects and
to find reliably as many groups as possible to track the
supercluster network.
}
{
We take into account various selection effects
due to the use of a magnitude limited sample. To determine
linking length scaling we study the
luminosity-density relation in observed groups. We follow the
changes in group sizes and mean galaxy number densities  within
groups when shifting nearby groups to larger distances.
As a result we show that the linking length should be a slowly
growing function with distance.
Our final sample contains 17143 groups in the equatorial,
and 33219 groups in the northern part of the DR5 survey
with membership $N_g \geq $ 2.
The group catalogue is available at our web-site
(\texttt{http://www.obs.ee/$\sim$erik/index.html}).
}
{
Due to a narrow magnitude window in the SDSS the group
catalogue based on this survey has been obtained  by moderately
growing linking length scaling law up to redshift z = 0.12.
Above this redshift the scaling law turns down.
In the redshift range z=0.12 - 0.2 only the cores are detected.
Along with applying weights when calculating luminosities it is
possible to use groups for determination of the large-scale
luminosity-density field.
}


\maketitle

\section{Introduction}

Groups and clusters of galaxies represent important ingredients in the
Universe for many purposes,  for example,  to test the large-scale structure or
the underlying cosmological model. The cluster catalogues by Abell
(\cite{abell}) and Abell \etal\ (\cite{aco}) were constructed by visual
inspection of Palomar plates.   The catalogues of the new generation of galaxy
groups were the Las Campanas catalogue of groups by Tucker \etal
(\cite{Tucker00}),  the catalogues based on SDSS (Sloan Digital Sky Survey) data
releases (EDR,  DR1,  DR2,  DR3,  DR4,  DR5) and the 2dFGRS (2 degree Field Galaxy
Redshift Survey) data releases (100K,  final,  Colless \etal
\cite{col01}, \cite{col03}).   This inspired numerous research teams to investigate 
more refined cluster finding algorithms and to compile catalogues of galaxy
systems (de Propris \etal\ \cite{dep02a},  Merchan \& Zandivarez \cite{mer02}, 
\cite{mer05},  Bahcall \etal\ \cite{bac03},  Lee \etal\ \cite{lee04},  Eke \etal
\cite{eke04},  Yang \etal\ \cite{yang05},  Einasto \etal\ \cite{ein05},  Goto
\cite{goto02},  Weinmann \etal\ \cite{wein06},  Tago \etal\ \cite{tago06},  Berlind
\etal\ \cite{ber06}). 

In our previous paper Tago \etal\ (\cite{tago06},  hereafter Paper~1) we have extracted 2dFGRS 
groups, and we have
given an extensive review of papers dedicated to group search methods and to
published group catalogues.  In this introduction we present a short review of
studies of galaxy groups. 

In recent years a number of new group finding algorithms and modified well
known methods have been applied (Goto \etal\ \cite{goto02},  Kim \etal
\cite{kim02},  Bahcall \etal\ \cite{bac03},  review by Nichol \cite{nic04}, 
Koester \etal\ \cite{koe07}).   However,  the friends-of-friends method (FoF, 
sometimes called percolation method) remains the most frequently applied for
redshift surveys.

\begin{table*}
      \caption[]{The SDSS DR5 Main samples used,  and the FoF parameters
      for the group catalogue (DR4 is for comparison but not studied)}
         \label{Tab1}
         \begin{tabular}{cccccccccc}
            \hline\hline
            \noalign{\smallskip}
          Sample &  $RA, \lambda$ & $DEC, \eta$ & $N_{gal}$ &
        $N_{groups}$&$N_{single}$&$\Delta V_0$ & $\Delta R_0$ &
    $ z_* $ & $ a $   \\
    & deg & deg &  & & & km/s &  Mpc/h &  & \\
            \noalign{\smallskip}
            \hline
          1 & 2 & 3 & 4  & 5 &
          6 & 7 & 8 & 9 & 10 \\
            \hline
            \noalign{\smallskip}

SDSS DR4 E  & 120... 255 & -1... 16     &  116471  & 16244  & 65016 &
250 & 0.25 & 0.138 & 1.46 \\

SDSS DR4 N  & -63... +63  & 6... 39 & 197481  & 25987   & 115488 &
250 & 0.25 & 0.138 & 1.46 \\

\\
SDSS DR5 E  & 120... 255 & -1... 16     &  129985  & 17143  &  75788 &
250 & 0.25 & 0.055 & 0.83 \\

SDSS DR5 N  & -63... +63  & 6... 39 & 257078  & 33219   & 152234 &
250 & 0.25 & 0.055 & 0.83 \\

            \noalign{\smallskip}
            \hline
         \end{tabular}\\

\small\rm\noindent Columns:
\begin{itemize}
\item[1:] the subsample of the SDSS redshift catalogue used, 
\item[2:] right ascension limits for the equatorial (E) sample,  $\lambda$
  coordinate limits for the northern (N) sample (degrees), 
\item[3:] declination limits for the E sample,  $\eta$ coordinate limits for
  the N sample (degrees), 
\item[4:] number of galaxies in a subsample, 
\item[5:] number of groups in a subsample, 
\item[6:] number of single galaxies, 
\item[7:] the FoF linking length in radial velocity,  for $z=0$, 
\item[8:] the FoF linking length in projected distance in the sky
  ,  for $z=0$, 
\item[9:] the characteristic scaling distance for the linking length
  ,  see Eq.~\ref{lz},  Sec.~5, 
\item[10:] the scaling amplitude for the linking length,  see
  Eq.~\ref{lz},  Sec.~5. 
\end{itemize}
   \end{table*}

Recently several authors have compiled group catalogues using the 2dF
Galaxy Redshift Survey.  One of the largest sample of groups has been compiled
by Eke \etal\ (\cite{eke04}),  who compared the real group samples with
samples found for simulated 2dF redshift survey galaxies.  
Yang \etal\ (\cite{yang05}) applied more strict
criteria in group selection,  and as a result have obtained a 2dF group
catalogue that contains mainly compact groups and a larger fraction of
single galaxies.  In Paper~1  we applied  criteria yielding groups
of galaxies with statistical properties between these two catalogues. 

Using earlier releases of the SDSS Lee \etal\ (\cite{lee04}, EDR), 
Merchan and Zandivarez (\cite{mer05}, DR3),  Goto (\cite{goto05}, DR2),  Weinmann
\etal\ (\cite{wein06}, DR2,  see for details Yang \etal\ \cite{yang05}), 
Zandivarez \etal\ (\cite{zan06}, DR4),  Berlind \etal\ (\cite{ber06}, DR3) have
obtained catalogues of groups (and clusters) of galaxies with rather different
properties.   In the present paper we have applied a FoF group search method
for the recent public release (DR5) of the SDSS.   All these group
catalogues are constructed on the basis of spectroscopic data of galaxy
catalogues using certain selection criteria.  The most important data and
properties for these catalogues (if available) are presented in
Table~\ref{Tab2}. 

Apart from the other authors Berlind \etal\ (\cite{ber06}) have used
volume-limited samples of the SDSS.  This yielded one of the most detailed
search method and reliable group catalogue(s).   Recently Paz \etal\ 
(\cite{paz06}) studied shapes and masses of the 2dFGRS groups (2PIGG),  Sloan
Survey Data Release 3 groups and numerical simulations,  and founda strong
dependence on richness. 

Papers dedicated to group and cluster search show a wide range of both sample
selection as well as cluster search methods and parameters.  The choice of
these parameters depends on the goals of the group catalogues obtained.   In
Paper~1 we drew a conclusion that in previous group catalogues the
luminosity/density relation in groups have not been applied.  In this paper we
apply this property of the observed groups to create a group catalogue for an
extended sample of the SDSS DR5. 

Selection effects in data are important factors in choosing galaxy selection
methods and understanding group properties.   In the present paper we
investigate various selection effects in SDSS (described in details in
Paper~1) which influence compilation of group catalogues.   We applied for the
SDSS DR5 (the last published data release) the well-known friends-of-friends
(FoF) algorithm.  Considering earlier experiences we selected a series of
procedures discussed below.

The data used are described in Section 2.   Sect.  3 discusses the
group-finding algorithm.  Selection effects,  which influence the
choice of parameters for the FoF procedure are discussed in Sect.  4. 
To select an appropriate cluster-finding algorithm we analyse in Sect.  5
how the properties of groups change,  if they are observed at various
distances.   Section 6 describes the final procedure
used to select the groups,  and the group catalogue.  We also estimate
luminosities of groups; this is described in Section 7. 
In the last Section we compare our groups with groups found by
other investigators,  and present our conclusions.  
As in Paper~1 we use for simplicity the term ``group'' for all objects in our
catalogue including also rich clusters of galaxies.

\section{The Data}

In this paper we have used the data release 5 (DR5) of the SDSS
(Adelman-McCarthy \etal \cite{ade07}; see also \cite{ade06},DR4) 
that contains overall 674749 galaxies
with observed spectra.  The spectroscopic survey is complete  from 
${\rm r} =14.5$ up to ${\rm r} =17.77$ magnitude.

We have restricted our study with the main galaxy sample obtained from the
SDSS Data Archive Server (DAS) which reduced our sample down to 488725
galaxies.  In present status the survey consists of two main contiguous areas
(northern and equatorial, hereafter N and E samples, respectively), and 3
narrow stripes in the southern sky and a short stripe at high declination.  We
have excluded smaller areas from our group search.  For the two areas the
coordinate ranges are given in Table~\ref{Tab1}.

We put a lower redshift limit $z=0.009$ to our sample with the aim to exclude
galaxies of the Local Supercluster.  As the SDSS sample becomes very diluted
at large distances, we restrict our sample by a upper redshift limit $z=0.2$.
Later we see that for our purposes this SDSS main sample is more or less
homogeneous up to $z=0.12$.
 
\begin{figure}[ht]
\centering
%\resizebox{0.45\textwidth}{!}{\includegraphics*{duplicate.eps}}
\resizebox{0.5\textwidth}{!}{\includegraphics*{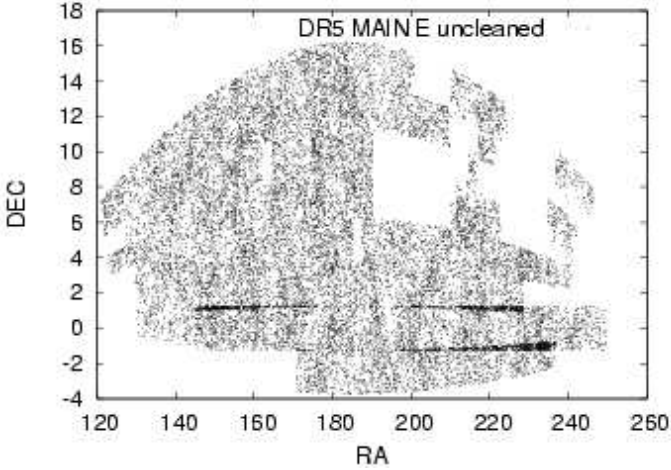}}
\caption{Duplicate galaxies in the sample E appearing as an increased density
of groups at the boundaries of the data releases 1 and 2. 
}
\label{fig:duplicate}
\end{figure}

\begin{figure}[ht]
\centering
\resizebox{0.5\textwidth}{!}{\includegraphics*{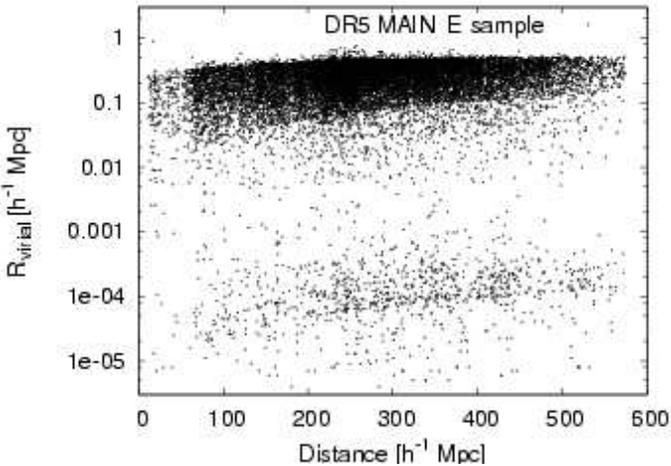}}
\caption{Duplicate galaxies in the sample E appearing as a separated
mode (due to false pairs at very low value of virial radius) in the 
virial radius - distance relation of groups. 
}
\label{fig:rvirduplicate}
\end{figure}

We have found duplicate galaxies due to repeated spectroscopy for a number of
galaxies in the DAS Main galaxy sample.   We have excluded from our sample those
 duplicate entries which have spectra of lower accuracy.   There were two
types of duplicate galaxies.  In one case duplicates had exactly identical ID
numbers, coordinates and magnitudes; they were simple to find out and to exclude. 
Another kind of duplicates had slightly different values of coordinates and
magnitudes.   This kind of duplicates cannot be seen in the sky distribution of
galaxies but were discovered as an enhanced number density of galaxy pairs
after the FoF procedure.  The majority of the second kind of duplicates have
been found at the common boundary of the data releases DR1 and DR2 (at DEC
$-1.25$ and $+1.25$).   We have excluded them as duplicate galaxies due to features
seen in Figure~\ref{fig:duplicate} and Figure~\ref{fig:rvirduplicate}.   In
total we have excluded from both samples 6439 identical galaxies and 1480
galaxies with slightly different data. 

The total number of galaxies has reduced to 129985 galaxies in the equatorial
sample and to 257078 galaxies in the northern sample.  Resulting data on the
samples are presented in Table~\ref{Tab1}.    In the present paper we have studied
only the SDSS DR5 release.  The redshifts were corrected for the motion
relative to the CMB.  For linear dimensions we use co-moving distances (see,
e.g., Mart{\`\i}nez \& Saar \cite{mar03}), computed with the standard cosmological
parameters: the Hubble parameter $H_0=100 h$, the matter density $\Omega_m =
0.3$, and the dark energy density $\Omega_{\Lambda} = 0.7$.

\section{Friends-of-friends algorithm}

One of the most conventional methods to search for groups of galaxies is
cluster analysis that was introduced in cosmology by Turner and Gott
(\cite{tg76}), and successfully nicknamed as the "friends-of-friends"
algorithm by Press and Davis (\cite{pd82}).  This algorithm along with the
percolation method started its world-wide use after suggestions by Zeldovich
\etal\ (\cite{zes82}) and by Huchra \& Geller (\cite{hg82}).  In Paper 1 we
have explained the FoF method and the role of linking length (or neighbourhood
radius) in detail. To summarize here in short: galaxies can be attributed to
systems using the FoF algorithm with a certain linking length.

Our experience and analysis show that the choice of the FoF parameters depends
on goals of the authors.  For example Weinmann \etal\ \cite{wein06} searched
for compact groups in a SDSS DR2 sample.  They applied strict criteria in FoF
method and obtained, as one of the results, a lower fraction of galaxies in

Berlind \etal\ (\cite{ber06} applied the FoF method to volume-limited 
samples of the SDSS (see Table~\ref{Tab2}).  Their goal
was to measure the group multiplicity function and to constrain dark halos.
The applied uniform group selection has reduced the incompleteness of the
sample, but it led also a lower number density of galaxies and of groups.

In this paper our goal is to obtain DR5 groups for a further determination of
luminosity density field and to derive properties of 
the network of the galaxy distribution.   Groups are mostly density
enhancements within filaments,  and rich clusters are high-density peaks of the
galaxy distribution in superclusters (Einasto et al.  \cite{einm03c}, 
\cite{einm03d},  \cite{ein07a},  \cite{ein07b}).  Hence,  our goal is to find out
as many groups as possible to track all of the supercluster network.   We
realize that differences in the purposes of the different papers which gives a fairly
wide range of group properties. 

A Virialisation condition, or a certain density contrast as alternative methods
do not work universally for all density ranges of galaxy distribution.
However, the similar problem arises in the case of FoF method.  As shown by
Einasto et al.  (\cite{e84}), it is not easy to find a suitable linking length
even for a volume-limited sample of galaxies.  The same conclusion has been
recently reached by Berlind \etal\ (\cite{ber06}), based on a much more larger
sample and a more detailed analysis.  The problem arises due to the variable
mean density of galaxies in different regions of space.  Additional
difficulties arise in case of flux-limited samples of galaxies if the linking
length depends also on the distance from the observer.  In the original
analysis by Huchra \& Geller the linking length was chosen as $l\sim f^{-1/3}$,
where $f$ is the selection function of galaxies.  This scaling corresponds to
the hypothesis that with increasing distance the galaxy field, and the groups,
are randomly diluted.  A recent summary of various methods to find clusters in
galaxy samples is given by Eke et al.  (\cite{eke04}).

\begin{figure}[ht]
\centering
\resizebox{0.45\textwidth}{!}{\includegraphics*{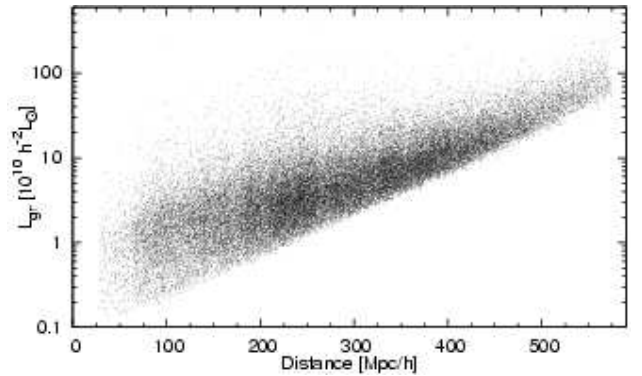}}
\caption{The total estimated luminosities for groups 
 as a function of distance from the observer. 
}
\label{fig:2}
\end{figure}

There exists a close correlation between luminosities of galaxies in groups
and their positions within groups: bright galaxies are concentrated close to
the center, and companions lie in the outskirts (for an early analysis of this
relationship see Einasto et al.  \cite{eskc74}, for a recent discussion see
Paper~1).  In Paper~1 we have found that while constructing group catalogues
in the 2dFGRS a slightly growing linking length with distance has to be used.

A similar problem arises in the SDSS.  As selection effects were
analyzed in detail in Paper~1, then we shall discuss only shortly the selection
effects in the SDSS survey.  We perform tests to find an optimal set of
parameters for the FoF method in this study.

\section{Selection effects}

\subsection{Selection effects in group catalogues}

Main selection effects in group catalogues are caused by the fixed interval of
apparent magnitudes in galaxy surveys (see for details in Paper~1).  This
effect is shown for SDSS DR5 groups in Fig.~\ref{fig:2}.

\begin{figure}[ht]
\centering
\resizebox{0.45\textwidth}{!}{\includegraphics*{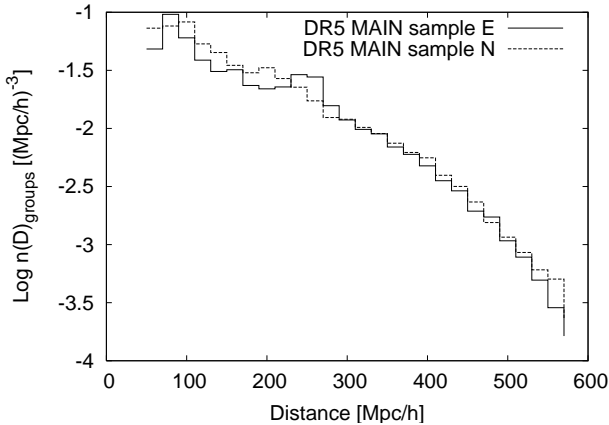}}
\caption{The number density of the SDSS DR5 MAIN E and N samples of 
groups in log scale as a function of distance from the observer . 
}
\label{fig:3}
\end{figure}

\begin{figure}[ht]
\centering
\resizebox{0.45\textwidth}{!}{\includegraphics*{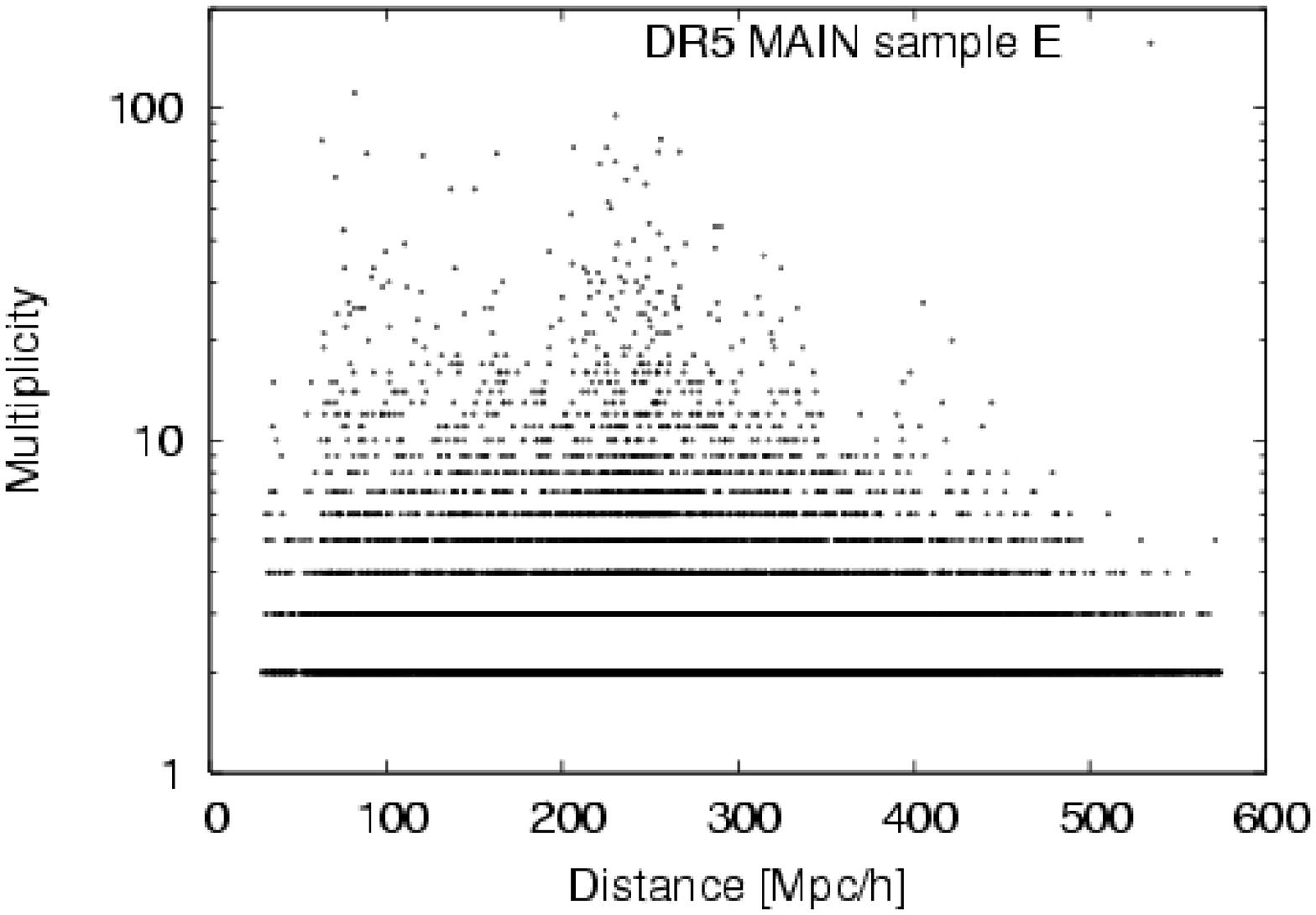}}
\caption{The multiplicity of groups of the sample E 
 as a function of distance from the observer. 
}
\label{fig:Nrich}
\end{figure}

\begin{figure}[ht]
\centering
\resizebox{0.45\textwidth}{!}{\includegraphics*{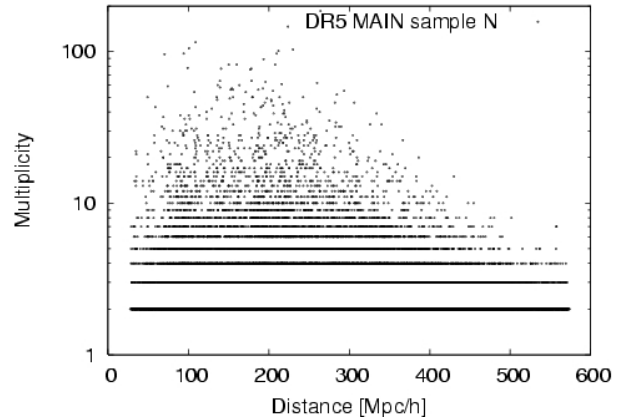}}\\
\caption{The multiplicity of groups of the sample N
 as a function of distance from the observer. 
}
\label{fig:NrichN}
\end{figure}

The main consequence of this selection effect is the inhomogeneous spatial
distribution of groups: the decrease of the volume density of groups with
increasing distance.  The mean volume density of groups as a function of
distance is plotted in Fig.~\ref{fig:3}, separately for the northern and the
equatorial area.

A consequence of this effect is richness (multiplicity) of groups as a
function of redshift.  In Figs.~\ref{fig:Nrich} and \ref{fig:NrichN} we show the
multiplicity of groups (the number of member galaxies) as a function of
distance from the observer for the E and N samples,  respectively.   We see that
rich groups are seen only up to a distance of about 300~\Mpc,  thereafter the
mean multiplicity decreases considerably with distance.   This selection effect
must be accounted for in the multiplicity analysis.   

\subsection{Selection effects in group sizes}

Sizes of groups depend directly on the choice of the linking length, or more
generally on its scaling law.  Strong selection effects can be observed
here, also.  As an example, the median sizes of the distant 2PIGG groups (Eke
\etal\ \cite{eke04}) are 7 times larger than those for the nearby groups.

\begin{figure}[ht]
\centering
\resizebox{0.45\textwidth}{!}{\includegraphics*{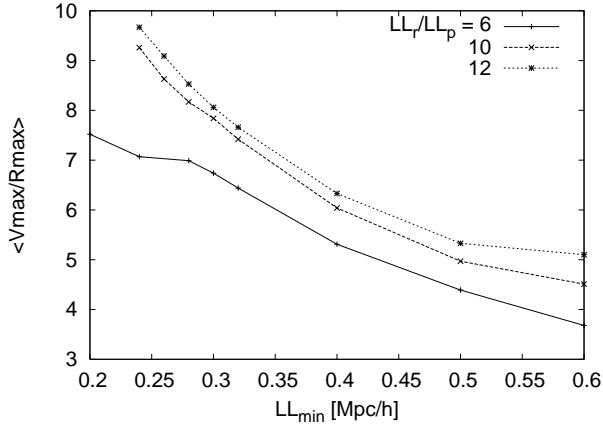}}
\caption{Mean ratio of radial and perpendicular sizes of groups
 in the sample E as a function of starting value of linking length
for three values of linking lengths ratios. 
}
\label{fig:VRmeanratio}
\end{figure}

Usually the ratio of radial and transversal linking lengths $ \Delta V_0 /
\Delta R_0$ is a constant in the FoF process of search of groups.  As noted by
Einasto \etal\ (\cite{e84}),  and Berlind \etal\ (\cite{ber06}) it is impossible
to fulfill all requirements with any combination of these linking lengths.   We
try to find the ratio $ \Delta V_0 / \Delta R_0$ which is the best to fulfill
the size ratio of observed groups which was determined by other studies. 
Figure~\ref{fig:VRmeanratio} demonstrates how the mean group size ratio
depends on initial linking length (LL) for three different $\Delta V / \Delta R $ ratio: 6,  10, 
and 12.  If we accept from other considerations the initial $\Delta R_0  =
0.25$ \Mpc, then we could find the best ratio $ \Delta V_0 / \Delta R_0$ to be 10
( at $\Delta R_0=0.25$ the curve 10 is the closest to the same value of
mean size ratio). 

On the other side,  if we accept size ratio 10 (for example from detailedd study
of cluster shape in redshift space) we could conclude the best $\Delta R_0$ to
be 0.25~\Mpc\ where the curve $<V/R>(\Delta R_0)$ reach the size ratio $\Delta V
/ \Delta R= 10$ in Figure~\ref{fig:VRmeanratio}. 

It is difficult to reliably model the galaxy populations in DM-haloes.  Here
we summarize in short a solution of the problem.

At large distances from the observer, only the brightest cluster members are
visible, and
these brightest members form compact cores of clusters, with sizes much less
than the true size of the clusters.
This effect work in the opposite direction to the increase of the linking
length, and it might cancel it out.
Next we describe the empirical scaling of
the linking length by shifting of the observed groups to growing distances.

\section{Scaling of linking length}

In the majority of papers dedicated to group search authors, the group finders
are tuned using mock $N$-body catalogues (e.g.  Eke \etal\ \cite{eke04}; Yang
\etal\ \cite{yang05}).  The mock group catalogues are homogeneous and all
parameters of the mock groups can be easily found and applied for search of
real groups.  Still mock groups are only an approximation to the real groups
using model galaxies in dark matter haloes.  As we have noted, it is difficult
to properly model the luminosity-density correlation found in real groups.

Starting from these considerations we have used observed groups to study the
scaling of group properties with distance.  The group shifting procedure is
described in detail in Paper~1.  As this is an important part of our search
method, then we present here the method i short and present the results for the SDSS
DR5 groups.
 
We created test group catalogues for the sample SDSS DR5 E with constant and
variable linking lengths, selected in the nearby volume $d < 100$~\Mpc all
rich groups (with multiplicity $N_{gal}\ge 20$, in total 222 groups).
Assuming that the group members are all at the mean distance of the group we
determined their absolute magnitudes and peculiar radial velocities.  Then we shifted
the groups step by step to larger distances (using a $z=0.001$ step in
redshift), and calculated new $k$-corrections and apparent magnitudes for the
group members.  As with increasing distance more and more fainter members of
groups fall outside the observational window of apparent magnitudes, the group
membership changes.  
We found new properties of the groups -- their
multiplicities, characteristic sizes, velocity dispersions and densities.  
We also calculated the minimum FoF linking length, necessary to keep the group
together at this distance.  

To determine that, we built the minimal spanning
tree for the group (see, e.g., Martinez and Saar \cite{mar03}), and found the
maximum length of the MST links.

\begin{figure*}[ht]
\centering
\resizebox{0.48\textwidth}{!}{\includegraphics*{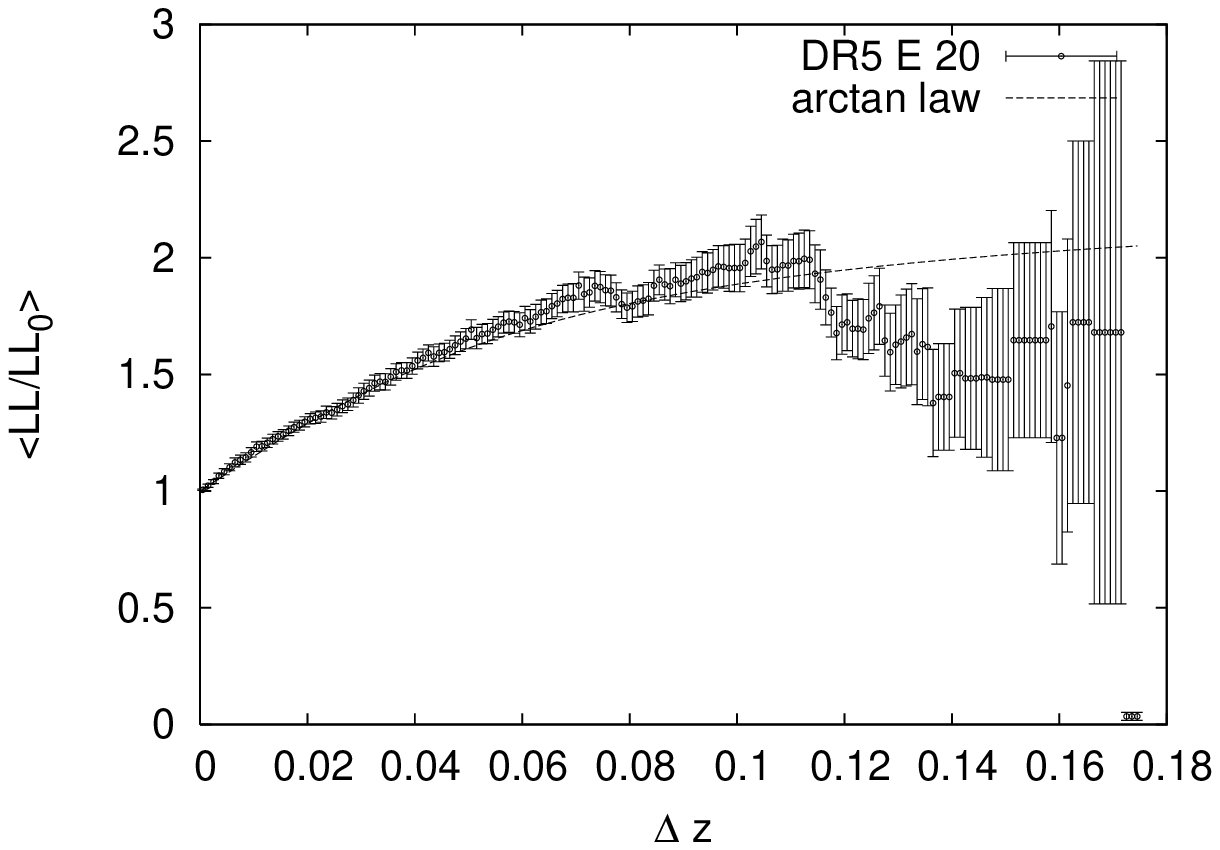}}
\hspace*{2mm}
\resizebox{0.48\textwidth}{!}{\includegraphics*{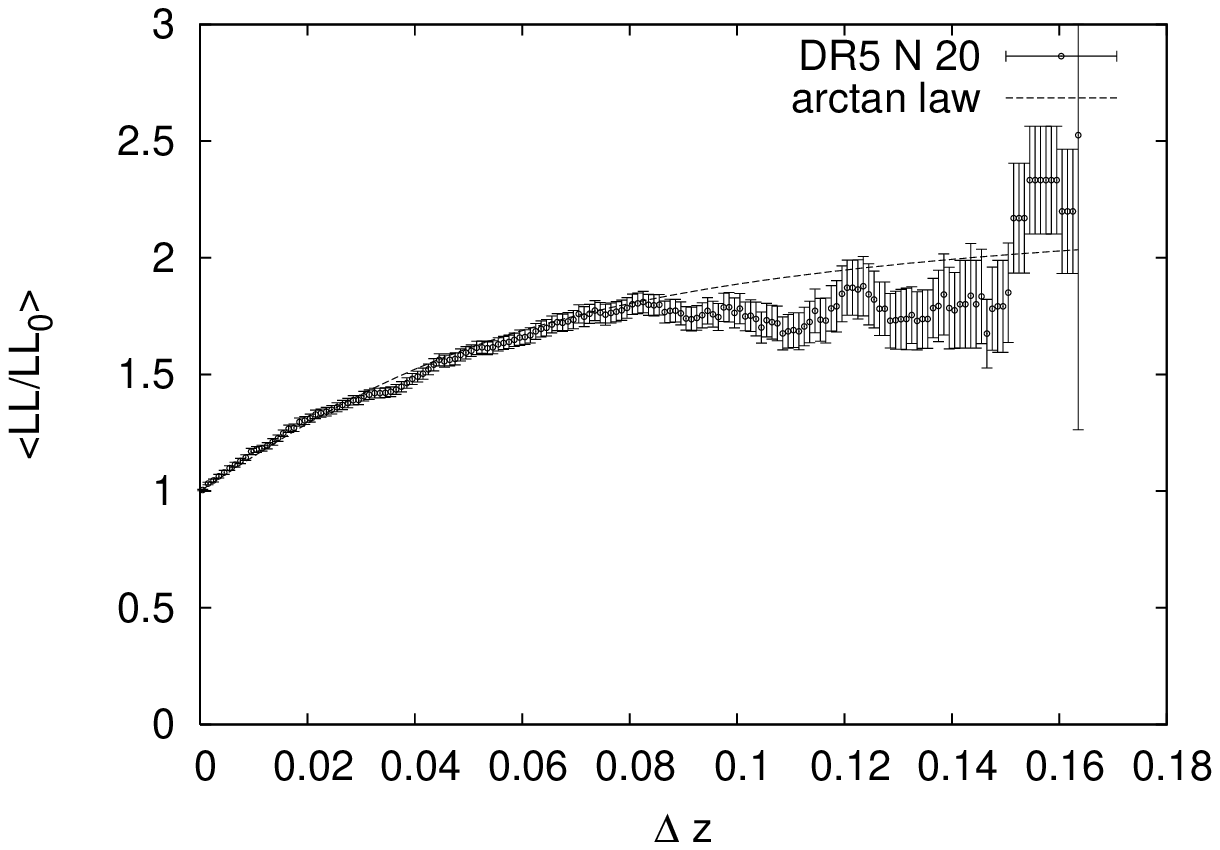}}
\caption{The scaling of the group FoF linking length with redshift
  for the samples DR5 E (left panel) and DR5 N (right panel).  The ordinate is
  the ratio of the minimal linking length $LL$ at a redshift $z$,  necessary to
  keep the group together,  to the original linking length $LL_0$ that defined
  the group at its initial redshift $z_0$; the abscissa is the redshift
  difference $\Delta z=z-z_0$. 
}
\label{fig:LLLawdr5}
\end{figure*}

As the original groups had different sizes and initial redshifts we found the
relative changes of their properties, with respect to the redshift change.
The individual linking length scaling paths have large scatter.  Therefore we
found the average scaling path from the individual paths.  In
Figure~\ref{fig:LLLawdr5} we present the main result of group shifting for our
linking length scaling law determination.

We fit the mean values of the linking lengths in $\Delta z=0.001$ redshift
bins (the step we used for shifting the groups).  We find our scaling law for
the case $n\ge 20$.  The fitting law is not sensitive to the richness of
groups involved in the LL scaling law determination.  The scaling law is
moderately different from the scaling law found for the 2dFGRS groups in
Paper~1 but still can be approximated by a slowly increasing arctan law.  
Due to narrow
magnitude window in SDSS, at higher values of $z$ only compact cores of
groups or binary galaxies have been found by FoF
method.  The deviation from the scaling
law corresponds to the redshift limit above which most groups discovered
correspond only to the compact cores of nearby groups.  Therefore, the
determination of the scaling law is a test for redshift limit of homogeneity
of the group catalogue. A good parametrization of the scaling low is
\begin{equation}
\label{lz}
LL/LL_0=1+a\, \mbox{arctan}(z/z_{\star}),
\end{equation}
where $a=0.83$ and $z_{\star}=0.055$. 

The main difference between the scaling laws of DR5 and 2dF groups is in the 
validity range.  This is due to different magnitude limits in these  
flux limited samples.  We consider this difference in more details below.   
The selection of initial groups should not influence much the scaling
of their properties with distance.  
We tested group search with three  different initial scaling laws for 
group selection
: two lengths constant and one varied with distance. 
The final scaling relation practically does not depend  on
the initial group selection (i.e.  on initial scaling law).

\section{Group catalogue}

\subsection{The group finder}

We adopt the scaling of the linking length found above, but we have to select
yet the initial values for the linking length.  In practice, only groups with
the observed membership $N_{gal} \geq 2$ are included in group catalogues.

\begin{figure*}[ht]
\centering
\resizebox{0.45\textwidth}{!}{\includegraphics*{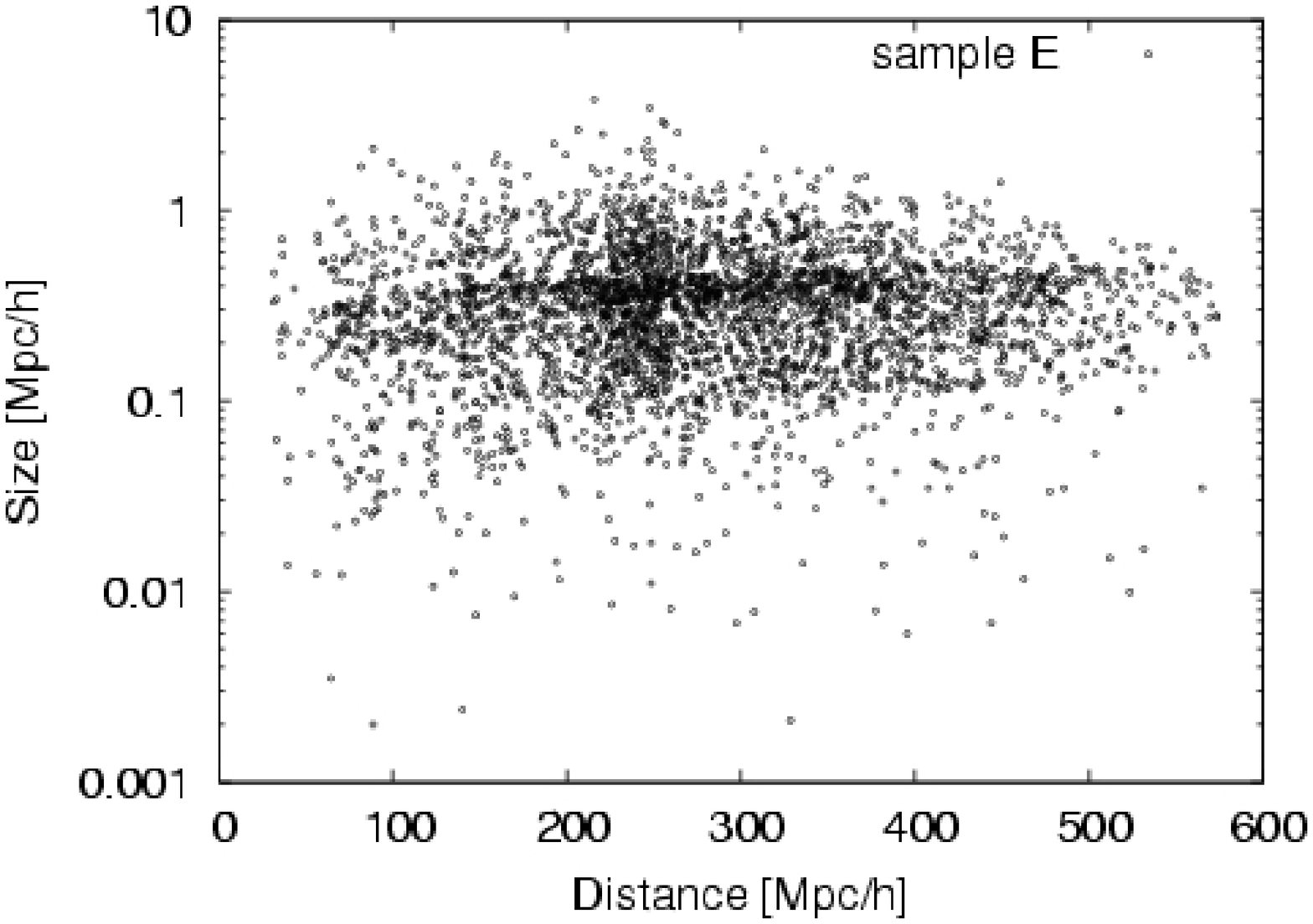}}
\hspace*{2mm}
\resizebox{0.45\textwidth}{!}{\includegraphics*{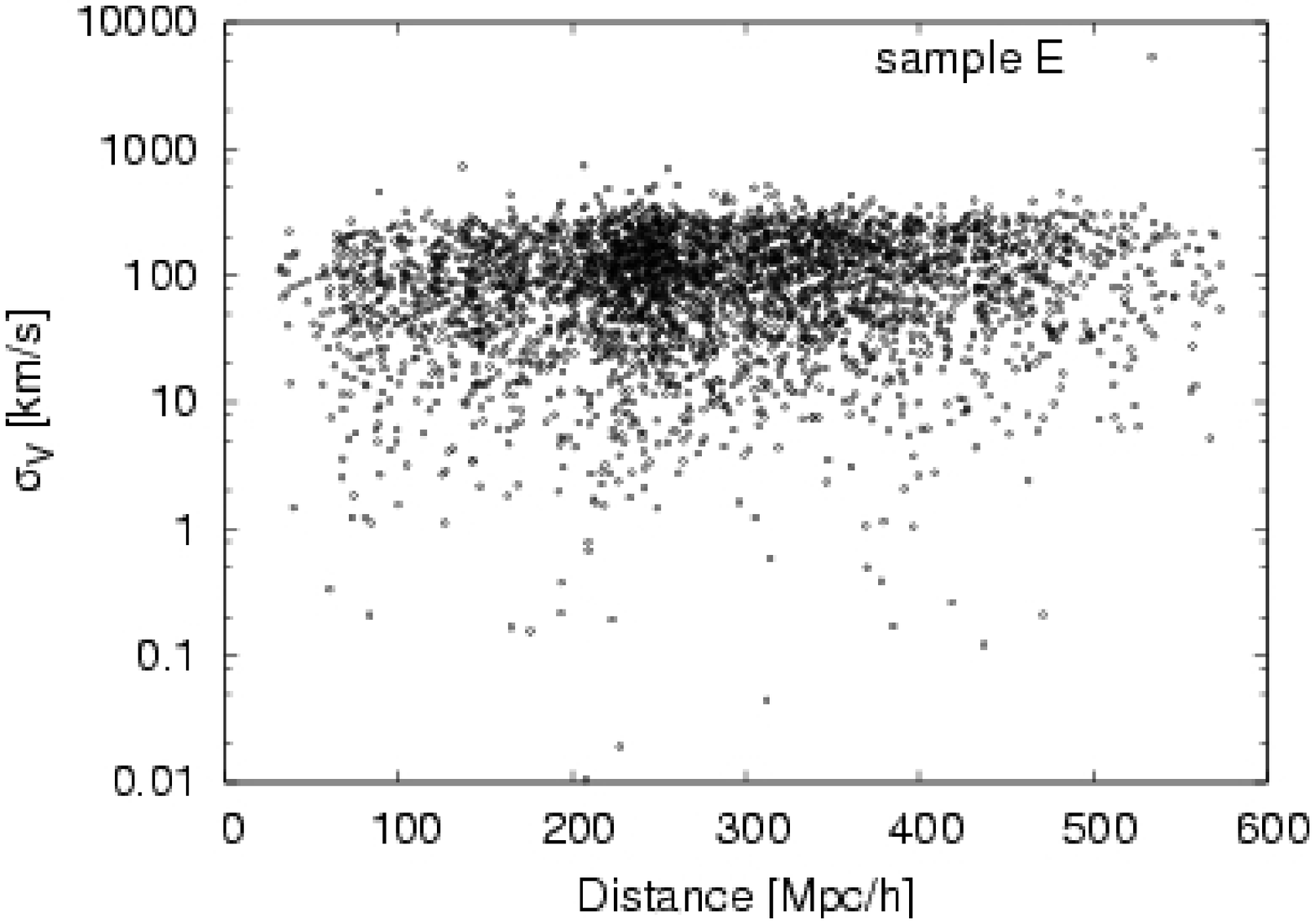}}\\
\caption{ Left panel : the (maximum projected) sizes of our SDSS DR5 groups
  in E sample as a function of distance. 
  Right panel shows the velocity dispersions in groups as a function of
  distance in the sample E.  The FoF parameters are  given in Table~\ref{Tab1}. 
}
\label{fig:10}
\end{figure*}

\begin{figure*}[ht]
\centering
\resizebox{0.45\textwidth}{!}{\includegraphics*{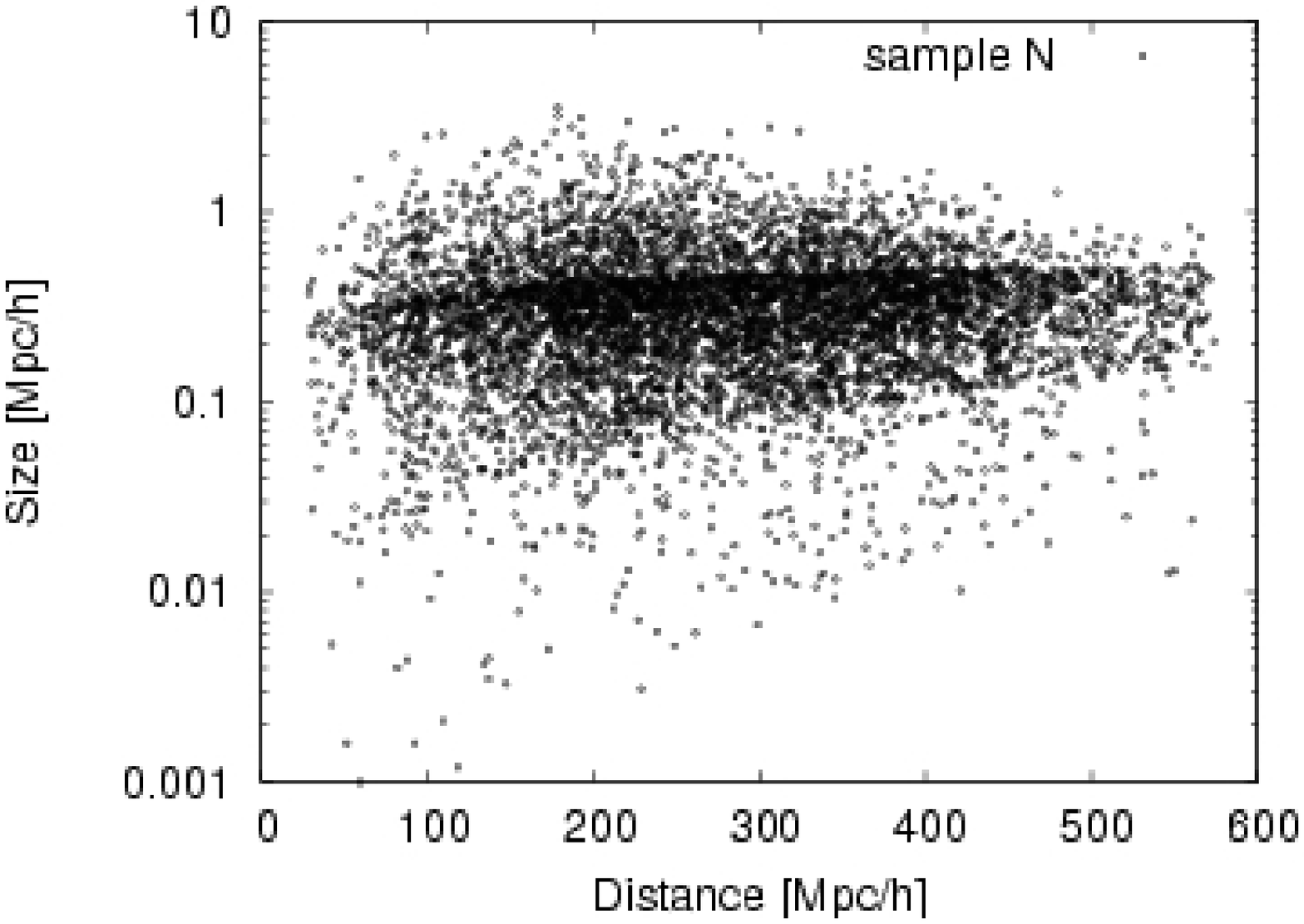}}
\hspace*{2mm}
\resizebox{0.45\textwidth}{!}{\includegraphics*{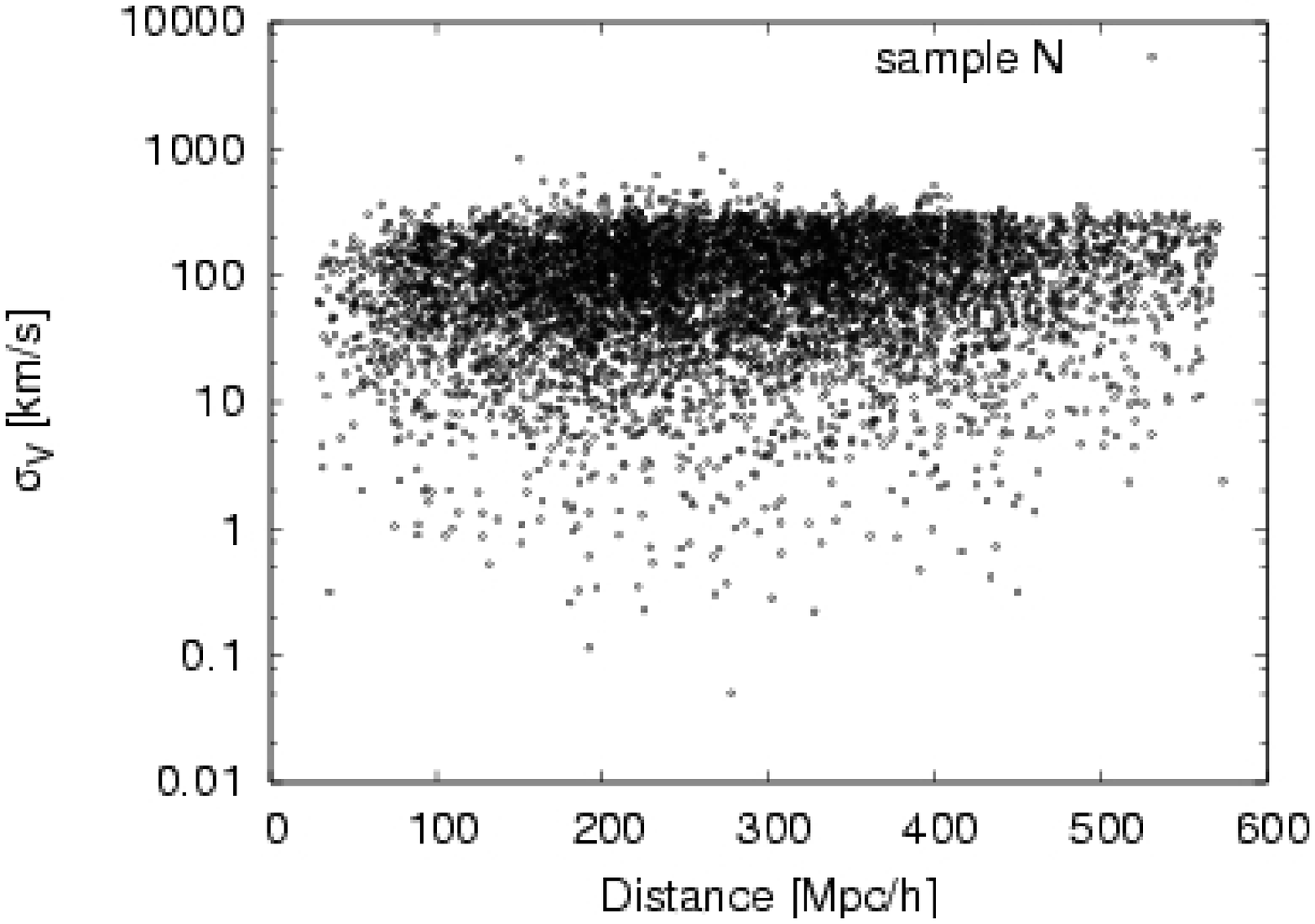}}\\
\caption{ Left panel : the (maximum projected) sizes of our SDSS DR5 groups
  in N sample as a function of distance. 
  Right panel shows the velocity dispersions of groups as a function of
  distance in the sample N. 
  The FoF parameters are  given in Table~\ref{Tab1}. 
}
\label{fig:sizesig}
\end{figure*}

In order to find the best initial linking lengths in the radial direction, we
tried a number of different parameter values, $\Delta V = 100-700$ km/s and
$\Delta R = 0.16 - 0.70$ \Mpc, and we chose finally the values which were
discussed above, and presented in Table~\ref{Tab1}.  Higher values for $\Delta
R$ leads to inclusion of galaxies from neighbouring groups and filaments.
Lower values for $\Delta V$ exclude the fastest members in intermediate
richness groups.

However, closer inspection show that one rich group has a richness much larger
($N=569$) than the rest of them.  This is the well-known nearby ($d=27$ \Mpc)
binary Abell cluster A2197/2199.  We consider this cluster as an exception,
and do not use lower LLs.  At slightly lower value of LL this cluster
fall apart and become the cluster with usual properties.

In Fig.~\ref{fig:10} we show the sizes of our groups of the final catalogue.
We define the size of the group as its maximum projected diameter, the largest
projected galaxy pair distance within the group.  We see that the sizes of
largest groups slightly increase with distance up to $d = 250$~\Mpc, and
thereafter slowly decrease.  This decrease is expected since in more distant
groups only bright galaxies are seen, and they form the compact cores of
groups.
The numbers of the groups and the FoF parameters (separately
for both SDSS DR5 regions) are given in Table~\ref{Tab1}.

\subsection{The final catalogue}

Our final catalogue (Table~\ref{Tab1}) includes 17143 groups in equatorial
area and 33219 groups in high declination area with richness $\geq 2$.  As an
example we present here the first lines of our group table (Table~\ref{Tab3}),
which include the following columns for each group:

\begin{itemize}
 \item[1)] group identification number;
 \item[2)] group richness (number of member galaxies);
 \item[3)] RA (J2000.0) in degrees (mean of member galaxies);
 \item[4)] DEC (J2000.0) in degrees (mean of member galaxies);
 \item[5)] group distance in \Mpc\ (mean comoving distance for member galaxies corrected
for CMB);
 \item[6)] the maximum projected size (in \Mpc);
 \item[7)] the rms radial velocity ($\sigma_V$,  in km/s);
 \item[8)] the virial radius in \Mpc\ (the projected harmonic mean);
 \item[9)] the luminosity of the cluster main galaxy (in units of $10^{10} h^{-2}
 L_{\sun}$);
 \item[10)] the total observed luminosity of visible galaxies ($10^{10} h^{-2}
 L_{\sun}$);
 \item[11)] the estimated total luminosity of the group  ($10^{10} h^{-2} L_{\sun}$). 
\end{itemize}

\begin{table*}
      \caption[]{First rows as an example of groups in the SDSS DR5 main
        galaxy catalogue   
 described in the present paper}
         \label{Tab3}
         \begin{tabular}{ccccccccccc}
            \hline\hline
            \noalign{\smallskip}
         $ ID_{gr}$ & $N_{g}$ &  $RA$ & $DEC$ & Dist &
        $Size_{sky}$&$\sigma_{V}$&$R_{vir} $ & $ L_{main}$ &
    $ L_{obs} $ & $L_{est}$   \\
           &  & [deg] & [deg] & [Mpc/h] & [Mpc/h] & [km/s] &  [Mpc/h] &
            [$10^{10} h^{-2} 
L_{\sun}]$& [$10^{10} h^{-2} L_{\sun} $] & [$ 10^{10} h^{-2} L_{\sun} $ ] \\
            \noalign{\smallskip}
            \hline
          1 & 2 & 3 & 4  & 5 &
        6 & 7 & 8 & 9 & 10 & 11 \\
            \hline
            \noalign{\smallskip}

     1  &    4 & 146.57633972 & -0.83209175 & 195.056 &  0.6823 &  53.7783 &
 0.33341 & 0.17353E+01 & 0.40818E+01 & 0.52815E+01 \\
     2  &    2 & 146.91120911 & -0.31007549 & 385.390 &  0.1291 &  25.2219 &
 0.12908 & 0.21835E+01& 0.41985E+01 & 0.10160E+02 \\
     3 &     3 & 146.88099670 & -0.49802899 & 249.334 &  0.1522 & 101.6915 &
 0.09505 & 0.27161E+01& 0.36896E+01  & 0.53377E+01 \\
     4   &   2 & 146.78494263 &  0.02115750 & 368.779 &  0.3185 & 173.4426 & 
 0.31840 & 0.37278E+01& 0.56619E+01 & 0.13310E+02  \\
     5   &   4 & 146.74797058 & -0.25555125 & 383.818 &  0.3404 & 191.9961 &
 0.15149 & 0.37084E+01& 0.99677E+01 & 0.24499E+02  \\

            \noalign{\smallskip}
            \hline
         \end{tabular}\\
   \end{table*}

The identification number is attached to groups by the group finder in
the order the groups are found. The calculation of luminosities is
described in the next section. 

We also give (in an electronic form) a catalogue of all individual
galaxies along with their group identification number and the group richness, 
ordered by the group identification number,  to facilitate search.   The
tables of galaxies end with a list of isolated galaxies (small
groups with only one bright galaxy within the observational window of
magnitudes); their group identification number is 0 and group richness
is 1.   All tables can be found at
\texttt{http://www.obs.ee/$\sim$erik/index.html}.   

\section{Luminosities of groups}

The limiting apparent magnitude of the complete sample of the SDSS catalog in
${\rm r}$ band is 17.77.  The faint limit actually fluctuates from field to
field, but in the present context we shall ignore that; we shall take these
fluctuations into account in our paper on the group luminosity function, based
on our 2dFGRS group catalogue (Einasto et al. \cite{ets07}).

We regard every galaxy as a visible member of a group or cluster within the
visible range of absolute magnitudes, $M_1$ and $M_2$, corresponding to the
observational window of apparent magnitudes at the distance of the galaxy.  To
calculate total luminosities of groups we have to find for all galaxies of the
sample the estimated total luminosity per one visible galaxy, taking into
account galaxies outside of the visibility window.  This estimated total
luminosity was calculated as follows (Einasto \etal\ \cite{e03b})
\begin{equation}
L_{tot} = L_{obs} W_L, 
\label{eq:ldens}
\end{equation}
where $L_{obs}=L_{\odot }10^{0.4\times (M_{\odot }-M)}$ is the
luminosity of a visible galaxy of an absolute magnitude $M$,  and
\begin{equation}
W_L =  {\frac{\int_0^\infty L \phi
(L)dL}{\int_{L_1}^{L_2} L \phi (L)dL}}
\label{eq:weight2}
\end{equation}
is the luminous-density weight (the ratio of the expected total luminosity to
the expected luminosity in the visibility window).   In the last equation
$L_i=L_{\odot} 10^{0.4\times (M_{\odot }-M_i)}$ are the luminosity limits of
the observational window,  corresponding to the absolute magnitude limits of
the window $M_i$,  and $M_{\odot }$ is the absolute magnitude of the Sun.   In
calculation of weights we assumed that galaxy luminosities are distributed
according to a two power-law function used by Christensen
(\cite{chr75}),  Kiang (\cite{kiang76}),  Abell (\cite{abell77}) and Mottmann \&
Abell (\cite{ma77})

\begin{equation}
 \phi(L)dL \propto (L/L^*)^\alpha(1+(L/L^*)^\gamma)^{(\delta / \gamma)}d(L/L^*) ,
\label{eq:twoplaw}
\end{equation}
where $\alpha $,  $\gamma$,  $\delta$ and $L^{*}$ are parameters.  We use two
power-law rather than Schechter function,  because it has
more freedom and it gives a better fit for the galaxy luminosity function. 

We used two power-law function with parameters: $\alpha = -1.123$,  $\gamma=
1.062$,  $\delta = -17.37$,  $L^{*} = 19.61$.  We have used all galaxies
(galaxies in groups and isolated galaxies) for finding the luminosity
function.  More detailed explanation about two power-law function and how we
derive the parameters are given in our paper on the 2dFGRS luminosity function
(Einasto et al.  \cite{ets07}). 

We derived $k$-correction for SDSS galaxies using the KCORRECT algorithm
(Blanton \& Roweis \cite{bla06}).   We also accepted $M_{\odot} = 4.52$ in the
${\rm r}$ photometric system. 

 We calculated for each group the total observed and corrected luminosities, 
and the mean weight
\begin{equation}
W_m = {\frac{\sum L_{tot, i}} {{\sum L_{obs, i}}}}, 
\label{eq:sum}
\end{equation}
where the subscript $i$ denotes values for individual observed galaxies in
the group,  and the sum includes all member galaxies of the system. 

\begin{figure}[ht]
\centering
\resizebox{0.45\textwidth}{!}
{\includegraphics*{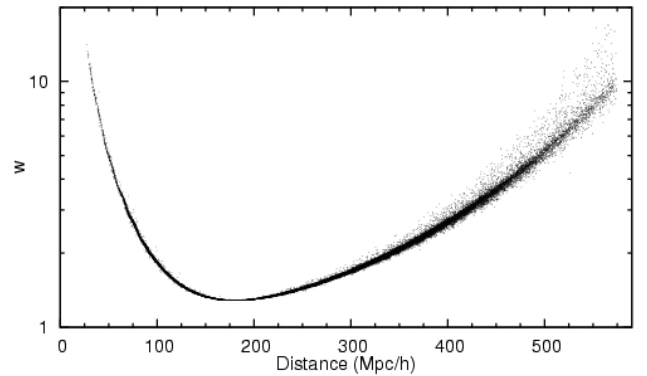}}
\caption{The mean weights of groups of the SDSS DR5 
  versus the distance from the observer. 
}
\label{fig:11}
\end{figure}

The mean weights for the groups of the SDSS DR5 are plotted as a function of
the distance $d$ from the observer in Fig.~\ref{fig:11}.  We see that the mean
weight is slightly higher than unity at a distance $d\sim 175$~\Mpc, and
increases both toward smaller and larger distances.  The increase at small
distances is due to the absence of very bright members of groups, which lie
outside the observational window, and at large distances the increase is
caused by the absence of faint galaxies.  The weights grow fast for very close
groups and for groups farther away than about 400~\Mpc.  At these distances
the correction factors start to dominate and the luminosities of groups become
uncertain.

In Fig.~\ref{fig:2} we show the estimated total luminosities of groups as a
function of distance.  We produced also colour figures that visualise the
luminosities of groups.  These are too detailed to be presented here, and can be
found in our web pages.  These figures show that the brightest groups have
corrected total luminosities, which are, in the mean, independent of distance.
This shows that our calculation of total luminosities is correct.

{\scriptsize
\begin{table*}
      \caption[]{Data for group catalogues based on the SDSS}
         \label{Tab2}
\begin{center}
      \[
         \begin{tabular}{llcccccccc}
            \hline\hline
            \noalign{\smallskip}
             Authors & Release, 
            Sample & $N_{gal} $ &
        $N_{gr}(n \geq 2)$ & $N_{gr}(n \geq 4)$&$z_{lim}$ & 
     $\Delta V_0 $ & $\Delta R_0$ & \%  ($\geq$ 2)  & \% ($\geq 4$)\\
     & & & & & & km/s & Mpc/h & & \\
            \noalign{\smallskip}
            \hline
          1 & 2 & 3 & 4  & 5 &
        6 & 7 & 8 & 9 & 10 \\
            \noalign{\smallskip}
            \hline

Merchan 2005 & DR3 Main    &  300000 &  & 10864  & 0 - 0.3   & 200 & &    &  22 \\

Goto 2005 & DR2 SQL  & 259497  & 335  &   & 0.03-  &   1000 & 1.5 &   & 6 ($n
\geq$ 20) \\ 

Weinmann 2006 & DR2 Main VAGC & 184425  & 16012  & 3720  & 0.01 - 0.2 &
0.3$^1$ &  0.05$^1$ &  30 & 15 \\

Berlind 2006 & DR3 sam14 VAGC & 298729  & &  & & & & &  \\
             & vol.lim.  Mr20 & 57332    & & 4119$^3$ & 0.015-0.1    & 0.75 &
             0.14 &  56.3 & 37.2$^3$ \\
             & vol.lim.  Mr19 & 37938    & & 2696$^3$ & 0.015-0.068  & 0.75 &
             0.14 &  58.9 & 40.7$^3$ \\
             & vol.lim.  Mr18 & 18959    & & 1362$^3$ & 0.015-0.045  & 0.75 &
             0.14 &  60.0 & 42.2$^3$ \\

Tago 2007 & DR5 Main DAS   & 387063  & 50362 & 9454  & 0.009 - 0.2   & 250 & 0.25
&  41.1   & 23.4  \\
            \noalign{\smallskip}
            \hline
         \end{tabular}
      \]
\end{center}

\small\rm\noindent Columns:

\begin{itemize}
\item[1:] authors of group catalog,
\item[2:] sample and release number, 
\item[3:] number of galaxies, 
\item[4:] number of groups ($n \geq$  2), 
\item[5:] number of groups ($n \geq$ 4),  
\item[6:] redshift limits for sample galaxies, 
\item[7:] the FoF linking length in radial velocity,  for $z=0$, 
\item[8:] the FoF linking length in projected distance in the sky
  ,  for $z=0$, 
\item[9:] fraction of galaxies in groups ($n \geq$ 2), 
\item[10:] fraction of galaxies in groups ($n \geq$ 4). 
\end{itemize}

\small\rm\noindent Notes:

$^1$ for Weinmann \etal\ groups linking lengths are in the units of mean
galaxy separation; 
 
$^3$ for Berlind \etal\ groups richness $n \geq 3$
 
* for Berlind \etal\  apparent magnitude  limit was $r \leq  17.5$ ,  for the
  rest $r \leq  17.77$ 

* group-finders :  
   
  Merchan: FoF + mock catalog + iterative group re-centering + Schechter LF
  for LL scaling 

  Goto:  FoF  + group re-centering 

  Weinmann:  FoF + DM halo mock catalog + group re-centering 

  Berlind:  FoF + DM halo mock catalog

  Tago:  FoF + DM halo mock  + Dens/Lum relation in groups for LL scaling

   \end{table*}
}

\section{Discussion and conclusions}

\begin{figure}[ht]
\centering
\resizebox{0.45\textwidth}{!}{\includegraphics*{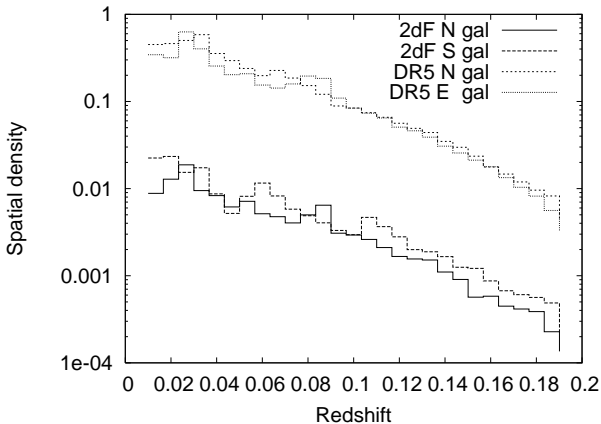}}
\caption{The number density of galaxies in the 2dF N and S samples,  and SDSS
  DR5 E and N samples  as a function of distance from the observer.  Histograms
  for 2dF are arbitrary shifted along  ordinate axis for clarity.   }
\label{fig:galden}
\end{figure}

\begin{figure}[ht]
  \centering \resizebox{0.5\textwidth}{!}  {\includegraphics*{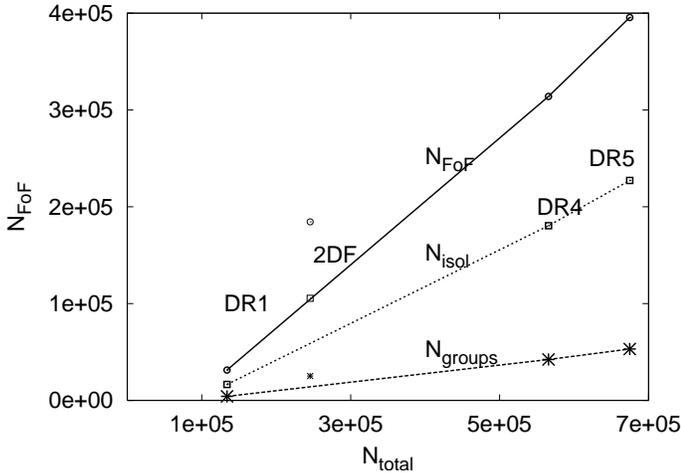}}
\caption{The number of sample galaxies,  groups and isolated galaxies
 involved in FoF procedure versus  total number of galaxies in releases of
 SDSS and 2dF surveys.  Note well defined proportional grows  with 
 releases of SDSS and a higher "yield" for 2dF.  These relations suggest that
 the FoF method has applied homogeneously to the different releases. 
}
\label{fig:dr-comp}
\end{figure}

\subsection{Some issues related to the poor de-blending}

Various potential caveats related to the automatic pipeline data reduction in
the SDSS have been discussed and flagged in the NYU-VAGC,  which is based on
the SDSS DR2 (Blanton \etal\ \cite{bla05}).  Most of these issues are related to
poor de-blending of large and/or of LSB galaxies with complicated morphology
(e.g. star-forming regions,  dust features etc.).  At low redshifts a number
of SDSS galaxies have been found shredded,  i.e.  a nearby large galaxy image
is split by target selection algorithm into several sub-images (e.g.  Panter
\etal\ \cite{panter07}).  Therefore,  the treatment of nearby galaxies requires special
care.  This potential bias is largely reduced in our new catalogue by means of
setting  reasonably high magnitude ($r > 14.5$) and redshift ($z > 0.009$)
limits,  which exclude most of luminous and/or nearby galaxies of the Local
Supercluster.

We have performed eyeball quality checks of a number of groups in the new
catalogue using the SDSS Sky Server Visual Tools.  We have inspected a) the
members of the 139 nearest ($z < 0.012$) groups -- 42 groups in the equatorial
(E) sample and 97 groups in the northern (N) sample; b) conspicuously dense
groups as evident on the bottom sections of the Figure~\ref{fig:rvirduplicate}, 
and of the Figures~\ref{fig:10} and \ref{fig:sizesig}.
 The results of these checks can be summarized as follows:
 
1) {\it De-blending errors.}  In the nearest 139 groups with initially 525
member galaxies poor de-blending has been noted for 21 (4\%) galaxies
distributed in 9 (6.5\%) groups.  Poor de-blending means either that the bright
galaxy is represented in the DR5 spectroscopic sample with a single off-center
source of typically reduced brightness,  or that the primary galaxy is shredded
into multiple (faint) \ion{H}{ii} regions. 

As an example of poor de-blending we refer to the group number 30644.
Its luminous member NGC 3995 ($B_T = 12.7$) with
knotty morphology is represented in the DR5 with 3 entries,  i.e.  with 3
distinct spectra of its \ion{H}{ii} knots of magnitudes $r$ = 12.6,  15.13,  
and 17.64, respectively.  Other three luminous group members NGC 3966
($m_B$ = 13.60), NGC 3994 ($B_T$ = 13.30), and NGC 3991 ($m_B$ = 13.50) 
are each represented in the DR5 by two knots with magnitudes
$r$ = 12.49,  16.88, and $r$ = 12.63,  16.60, and $r$ = 14.81, 17.89, respectively.
After excluding the knots with $r < 14.5$ those intrinsically luminous
galaxies will be represented in our catalogue by their faint(er) knots and
their true total magnitudes are underestimated by 1.5 - 3.5 magnitudes.
It appears to be one of the most severely biased nearby groups.

2) All the 25 very dense E groups with $R_{vir} < 1~ h^{-1}$ kpc,  distributed
in the bottom section of the Figure~\ref{fig:rvirduplicate},
 are results from duplicates.   Among them there
are 14 ''pairs'' ( i.e.  actually a single galaxy with two records in the DR5
spectroscopic sample),  7 "triplets" and 4 "quartets".  Among the N groups there
are only two duplicates in the given $R_{vir}$ range. 
  
3) Considering the Figures~\ref{fig:10} and 
\ref{fig:sizesig} (left panels) \\
-- all 13 groups with $Size < 1 h^{-1}$ kpc are among
those with $R_{vir} < 1~ h^{-1}$ kpc in the Figure~\ref{fig:rvirduplicate},
  i.e.  they are duplicates. \\
-- The conspicuous lower boundary of the tightly populated region (which varies
nearly proportional  to distance) is probably determined by the fiber collision
distance $\sim 55''$ of the survey. The groups distributed in the range
between this lower boundary and that of $Size = 10~ h^{-1}$ kpc are in the majority
real pairs,  i.e.  no duplicates. 
Pairs with $Size < 10 ~h^{-1}$ kpc 
are likely mergers, or advanced mergers
(with $1 < Size < 5 h^{-1}$ kpc).\\
 - The upper boundary of the tightly populated region likely results 
from the linking-length
scaling relation (1),  since there is no
single pair above this boundary.  That means,  our sample could be biased
against the wide (i.e.  in the majority optical) pairs. 
 
{\it To summarize:} As a result of our cursory checks we have found relatively
few bad de-blends,  either in form of mismatches between spectral targets and
optical centers,  or more severe shreddings of large and/or LSB galaxies. 
Although the redshifts are fine,  photometric and structural measurements are
often erroneous in such cases.  The fraction of groups checked so far is small, 
however it comprises the nearest,  i.e.  potentially most affected part of the
full sample.   We estimate that the net effect of de-blending errors will have
minor effect,  when working with large (sub)samples of groups.

\begin{figure}[ht]
\centering
\resizebox{0.5\textwidth}{!}
{\includegraphics*{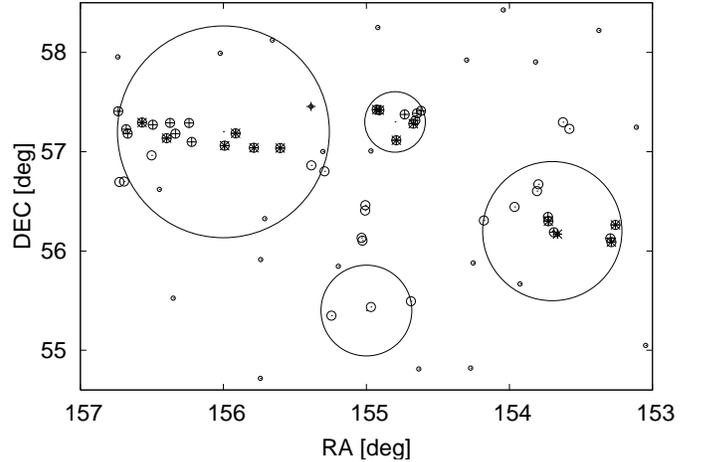}}
\caption{The eight nearby ($z < 0.04$) groups ($n \geq 2$) as identified 
  in this work in a relatively sparce filament.  The group members are shown
  with  circles  and four individual groups are encompassed with large circles. 
  The field galaxies in the same redshift range are marked with small circles. 
  For comparison,  the members of the corresponding Merchan \etal (\cite{mer05})
  groups ($n
  \geq 4$) are marked with tilted crosses ($\times$),  and those of the Berlind
  \etal (\cite{ber06}) groups (Mr18 sample,  $n \geq 3$) are shown with crosses.  Note
  that in Merchan \etal (\cite{mer05}) the rich,  elongated group is divided into two (NE
  and SW) subgroups,  which are nearly projecting to each other along the
  line-of-sight.   }
\label{fig:gr8344a}
\end{figure}

In Fig.~\ref{fig:gr8344a} we give an example of how the group-finder
algorithm works.  The comparison with groups Merchan \etal (\cite{mer05})
 and Berlind \etal (\cite{ber06})
shows that all three slightly different FoF algorithms identify quite similar
groups.  The criteria used in Merchan \etal (\cite{mer05}) tend to split the groups along the
line-of-sight and/or exclude the galaxies in outskirts of groups more easily.

\begin{figure}[ht]
\centering
\resizebox{0.50\textwidth}{!}{\includegraphics*{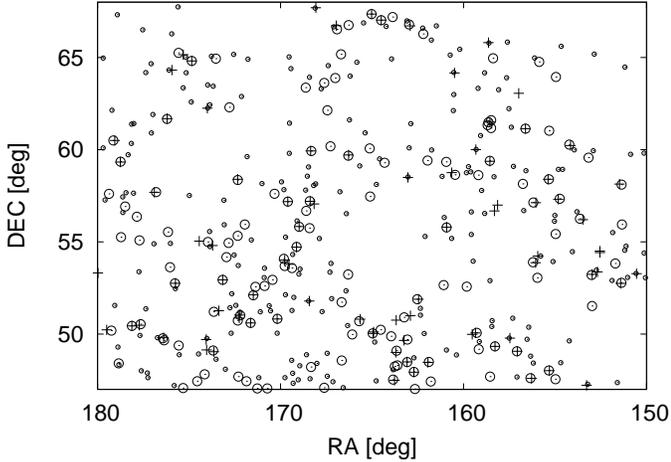}}
\caption{Groups by Berlind \etal (\cite{ber06}) Mr18 sample (crosses)
compared to our groups in the same redshift ($0.015 < z < 0.045$) and 
richness ($N_{gal} \geq 3$) range (large circles). The pairs
of galaxies ($N_{gal} = 2$) in our catalogue are shown with small circles. 
}
\label{fig:Berlind_dr5} 
\end{figure}

In Fig.~\ref{fig:Berlind_dr5} we compare the groups in the volume limited Mr18 sample 
of Berlind \etal (\cite{ber06}) to our groups in a similar redshift range. We conclude that we
can detect more groups (121 our groups versus 88 groups in Mr18) and slightly richer groups
(6.1 galaxies per one our group versus 5.5 galaxies in one Mr18 group), mainly due to 
inclusion of fainter ($Mr > -18$) galaxies.

\subsection{Comparison to other studies}

Earlier catalogues of the SDSS groups of galaxies, based on the first SDSS
releases, were obtained by Lee \etal (\cite{lee04}), Einasto \etal\ 
(\cite{e03b}).

At present there are five extensive catalogues of groups of galaxies available
to us which are obtained on the basis of the SDSS.  Although they are based on
different SDSS releases they have obtained by incremental addition of
new data to previous releases and observational method and parameters are the
same. We can reasonably compare these group catalogues.   Group catalogues
are different due to different group search parameters and not under-laying
samples of galaxies.  An important exception are 3 volume limited samples by
Berlind \etal  At the price of smaller galaxy sample they have the advantage that
the most serious incompleteness effect  of magnitude limited samples is
absent, the missing of faint galaxies in distant parts of the survey.               Some characteristics of the
catalogues are presented in Table~\ref{Tab2}.   An important characteristic to
compare the catalogues is the fraction of single (isolated) galaxies or
equivalently,  the fraction of galaxies in groups.   Single galaxies can be
considered as belonging to small groups or to haloes represented only by one
observed galaxy in the visibility window. 

Therefore,  we face the problem how to compare catalogues because different
group-finder criteria have been applied: richness and size of groups,  linking
lengths,  the ratio of los/perpendicular linking lengths,  etc.   These criteria
depend on the goals of a particular study.   The last two columns in the table
give the fraction of galaxies in groups of richness $n \geq 2$ and $n \geq 4$. 
These are 30 and 42 \% for the groups by Weinmann \etal\ and for our groups of
richness $\geq 2$,  and 22 and 18.3 \% for the groups by Merchan \etal,  and for
our groups of richness $\geq 4$,  respectively.   In fact, these values represent
the low richness end of the multiplicity function.   

We note that the fraction of galaxies in our 2dF GRS groups is very similar --
43 \% (Paper~1).  This suggest that the multiplicity distribution is a robust
characteristic being independent of these two surveys and  small
differences in initial parameters of FoF chosen.  We see that Weinmann's
groups which are intended to determine only compact groups, have remarkably
lower fraction of galaxies in groups  (30 \%) than ours. Comparing
these fractions for Merchan's and our groups the results are much closer (for
richness $n \geq$ 4).

Several studies have shown (see, e.g., Kim \etal\ \cite{kim02}) that different
methods give rather different groups for the SDSS sample.  The same is true
for the 2dFGRS groups (Paper~1).  Although catalogues cited in
Table~\ref{Tab2} are FoF-based, the results of Goto et al.  
(\cite{goto05}) have created a
cluster catalogue applying a very strong criteria for system search with a
purpose to study cluster galaxy evolution.  It is not much useful to compare
their catalogue with ours due to different purposes and the number of clusters.
However, we present for completeness also properties in
Table~\ref{Tab2}.  Weinmann \etal\ (\cite{wein06}) applied a more strict
criteria in group selection based on the idea that galaxies in a common dark
matter halo belong to one group.  As a result, they obtained a group catalogue
that contains mainly compact groups and a large fraction of single galaxies.

The most detailed search method and reliable group catalogue(s) have been
obtained by Berlind \etal\ (\cite{ber06}; SDSS collaboration).  Their purpose
was to construct groups of galaxies to test the dark matter halo occupation
distribution.  For this requirement to get highly reliable groups they choosed
a different way --- volume-limited samples of the SDSS.  This way has unwanted
result --- much smaller sample, but we see also (Table 2) the advantage ---
less incompleteness problems and a higher fraction of galaxies in groups than
in the other catalogues.  Berlind \etal\ (\cite{ber06}) demonstrated that
there exists no combination of radial and perpendicular linking lengths
satisfying all three important properties of groups (in mock catalogue):
the multiplicity function, the projected size and the velocity dispersion.

This could explain why the properties of group catalogues, presented in
Table~\ref{Tab2}, are so different.  We consider this fault as one of
justifications to use observed groups for determination of linking length
scaling law.

\subsection{Conclusions}

We have used the Sloan Digital Sky Survey Data Release 5 to create a new
catalogue of groups of galaxies.  Our main results are the following:

\begin{itemize}
  
\item[1)] We have taken into account selection effects caused by
  magnitude-limited galaxy samples.  Two most important effects are the
  decreasing of group volume density and the decreasing of the group richness
  with increasing distance from the observer.  We show that at large distances
  from the observer the population of more massive, luminous and greater
  groups/clusters dominates.  This increase of the mean size of groups is
  almost compensated by the absence of faint galaxies in the observed groups
  at large distances.  The remaining bright galaxies form a compact core of
  the group, this compensates for the increase of group sizes caused by
  domination of the population of more massive groups.  This confirms the
  similar luminosity/density relation found for 2dFGRS groups earlier.
  
\item[2)] We find the scaling of the group properties and that of the FoF
  linking length empirically,  shifting the observed groups to larger
  redshifts.  As the SDSS Main and 2dFGRS galaxies have similar redshift
  distributions and luminosity functions, then we find that the linking length
  scaling laws for these catalogues are very close, growing only slightly  by
  arctan law, but only up to the redshift $z=0.12$.   Beyond this redshift
  the scaling law decreases sharply.   At higher redshift we detect mainly compact
  cores of the groups due to more narrow magnitude range (visibility window)
  of the SDSS.  This scaling law method can be considered as a test to which
  redshift limit group-finder could be applied. 
  
\item[3)] We present a catalogue of groups of galaxies for the SDSS Data Release
  5.  We applied the FoF method with a slightly increasing linking length;
  the catalogue is available at the web page
  (\texttt{http://www.obs.ee/$\sim$erik/index.html}).
  
\item[4)]A wide variety of properties as a result of different purposes of the
  catalogues which involve different parametres for group search algorithms,
  and different samples.  
Others tried to establish parameters of the halo model
of the galaxy distribution. We provide a catalogue that was intented most
complete and representative for the survey volume. Thereby we best measure
 the large scale galaxy network over the survey volume.

\end{itemize}

\begin{acknowledgements}
  
  Funding for the Sloan Digital Sky Survey (SDSS) and SDSS-II has been
  provided by the Alfred P.  Sloan Foundation,  the Participating Institutions, 
  the National Science Foundation,  the U.S.  Department of Energy,  the National
  Aeronautics and Space Administration,  the Japanese Monbukagakusho,  and the
  Max Planck Society,  and the Higher Education Funding Council for England. 
  The SDSS Web site is http://www.sdss.org/. 
  
  The SDSS is managed by the Astrophysical Research Consortium (ARC) for the
  Participating Institutions.  The Participating Institutions are the American
  Museum of Natural History,  Astrophysical Institute Potsdam,  University of
  Basel,  University of Cambridge,  Case Western Reserve University,  The
  University of Chicago,  Drexel University,  Fermilab,  the Institute for
  Advanced Study,  the Japan Participation Group,  The Johns Hopkins University, 
  the Joint Institute for Nuclear Astrophysics,  the Kavli Institute for
  Particle Astrophysics and Cosmology,  the Korean Scientist Group,  the Chinese
  Academy of Sciences (LAMOST),  Los Alamos National Laboratory,  the
  Max-Planck-Institute for Astronomy (MPIA),  the Max-Planck-Institute for
  Astrophysics (MPA),  New Mexico State University,  Ohio State University, 
  University of Pittsburgh,  University of Portsmouth,  Princeton University, 
  the United States Naval Observatory,  and the University of Washington. 
  
  We are pleased to thank the SDSS collaboration for the DAS version of the
  fifth data release,  special thanks to James Annis.   We acknowledge the
  Estonian Science Foundation for support under grants No.  6104, 6106 and 7146,
  and the Estonian Ministry for Education and Science support by grant
  SF0062465s03.   This work has also been supported by the University of
  Valencia through a visiting professorship for Enn Saar and by the Spanish
  MCyT project AYA2003-08739-C02-01 (including FEDER).  J.E.   thanks
  Astrophysikalisches Institut Potsdam (using DFG-grant 436 EST 17/2/06),  and
  the Aspen Center for Physics for hospitality,  where part of this study was
  performed. 

\end{acknowledgements}

\end{document}